\documentclass[preprint,3p]{elsarticle}

\usepackage{lineno,hyperref}
\usepackage[usenam.pnges,dvipsnames,svgnames,table]{xcolor}
\modulolinenumbers[5]

\journal{Carbon}

\begin{document}

\title{Electronic, optical and thermal properties of highly stretchable 2D carbon Ene-yne graphyne}
%Optical, thermal and electronic properties of highly stretchable 2D carbon Ene-yne graphyne}
\author[buw]{Bohayra Mortazavi\corref{cor1}}
\ead{bohayra.mortazavi@gmail.com}

\author[bist]{Masoud Shahrokhi}

\author[tu]{Timon Rabczuk}

\author[ufrn]{Luiz Felipe C. Pereira\corref{cor1}}
\ead{pereira@fisica.ufrn.br}

\address[buw]{Institute of Structural Mechanics, Bauhaus-Universit\"at Weimar, Marienstr. 15, D-99423 Weimar, Germany}
\address[bist]{Institute of Chemical Research of Catalonia, ICIQ, The Barcelona Institute of Science and Technology, Av. Pa•sos Catalans 16, ES-43007 Tarragona, Spain}
\address[tu]{College of Civil Engineering, Department of Geotechnical Engineering, Tongji University, Shanghai, China}
\address[ufrn]{Departamento de F\'{\i}sica, Universidade Federal do Rio Grande do Norte, Natal, 59078-970, Brazil}

\cortext[cor1]{Corresponding authors}

\date{\today}

\begin{abstract}
Recently, a new carbon-based two-dimensional (2D) material, so called Òcarbon Ene-yneÓ (CEY), was successfully synthesized. In this work, we examine electronic, optical and thermal properties of this novel material. We studied the stretchability of CEY via density functional theory (DFT) calculations. Using the PBE and HSE06 functionals, as well as the G$_0$W$_0$ method and the Bethe-Salpeter equation, we systematically explored electronic and optical properties of 2D CEY. In particular, we investigated the change of band-gap and optical properties under uniaxial and biaxial strain. Ab-initio molecular dynamics simulations confirm that CEY is stable at temperatures as high as 1500 K. Using non-equilibrium molecular dynamics simulations, the thermal conductivity of CEY was predicted to be anisotropic and three orders of magnitude smaller than that of graphene. We found that in the visible range, the optical conductivity under high strain levels is larger than that of graphene. This enhancement in optical conductivity may allow CEY to be used in photovoltaic cells. Moreover, CEY shows anisotropic optical responses for x- and y- polarized light, which may be suitable as an optical linear polarizer. The comprehensive insight provided by the present investigation should serve as a guide for possible applications of CEY in nanodevices.
\end{abstract}

%\keywords{Graphene, thermal conductivity, low-dimensional materials, molecular dynamics}

\maketitle

\section{Introduction}

Two-dimensional (2D) materials are currently considered a new class of materials with numerous  applications. An exciting fact about the family of 2D materials is that it keeps expanding considerably since 2004, when the first mechanical exfoliation of graphene from graphite was accomplished \cite{Novoselov2004, Geim2007}. Currently, a wide variety of high quality 2D materials are available such as hexagonal boron-nitride \cite{Kubota2007, Song2010}, graphitic carbon nitride \cite{Thomas2008, Algara-Siller2014}, silicene \cite{Aufray2010, Vogt2012}, germanene \cite{Bianco2013}, transition metal dichalcogenides \cite{Geim2013, Wang2012a, Radisavljevic2011}, phosphorene \cite{Das2014, Li2014c}, holey C$_2$N sheets \cite{Mahmood2015, Mortazavi2016a,Tromer2017} and recently borophene \cite{Mannix2015}. 
Graphene in its defect-free and single-layer form exhibits a unique combination of ultra-high mechanical and thermal conduction properties, with an elastic modulus of ~1 TPa \cite{Lee2008}, tensile strength of ~130 $\pm$ 10 GPa \cite{Lee2008} and thermal conductivity in the range 3000-5000 W/m-K  \cite{Balandin2008,Ghosh2010,Balandin2011,Xu2014,Fan2017}, both at room temperature, outperforming all other known materials.
Nonetheless, in its pristine form graphene is a zero band-gap semiconductor, which limits its suitability for application in nanoelectronics. In fact, this limitation of graphene is one of the most appealing motivations to fabricate novel 2D materials with inherent semiconducting electronic properties. 

Synthesis of novel 2D materials with semiconducting character, made mainly from carbon atoms have been among the most interesting research topics in the field of novel  materials \cite{Baughman1987,Miller2016,Vipin2016,Koizumi2016,Wang2016d,Kang2016,Bizao2017}. New forms of carbon allotropes with sp or sp-sp$^2$ hybrid structures have already been investigated. For example, carbyne, a one dimensional chain of sp hybrid carbon atoms, which was first synthesized by Eastmond et al. \cite{Eastmond1972}, has attracted considerable attention. Carbyne presents metallic electronic response and superconductivity at room temperature \cite{Liu2013a, Cudazzo2010}. Up to date, the practical applications of linear carbyne have not been reported, because only low polymerization degree has been achieved, which has been far from the scale needed for realistic utilization. Graphyne \cite{ Baughman1987} is a class of 2D allotropes of carbon, which includes sp and sp$^2$ hybrid bonded carbon atoms arranged in a crystal lattice. Graphyne structures usually show a lattice of benzene rings connected by acetylene bonds. 
Recently, carbon Ene-yne (CEY), a novel full carbon 2D material was synthesized in multilayer form from tetraethynylethene by solvent-phase reaction \cite{Jia2017}. The crystal structure of CEY is similar to that of  graphyne, without the carbon benzene rings. Interestingly, the possibility of CEY existence was predicted first by Baughman et. al \cite{Baughman1987} in 1987, three decades earlier than its  experimental realization. In that first theoretical study, CEY was named 14, 14, 14-graphyne. Notably, CEY can yield a high specific capacity of 1326 mAh/g for lithium storage, which is three times higher than commercial graphite \cite{Jia2017}. The most recent experimental success in the fabrication of 2D CEY, highlights the importance of understanding its physical properties. In this way, evaluation of  mechanical,  electronic, optical and thermal properties of 2D CEY not only play a critical role in the design of novel nanodevices with CEY as a building block, but it may also find novel prospects and applications for their usage. Since the experimental characterization of  2D materials is rather challenging, complex and time consuming, computational  investigations can be considered as promising alternatives \cite{Mortazavi2017}. To this aim, in the present work we performed ab-initio calculations and extensive atomistic simulations to evaluate mechanical, electronic, optical and thermal properties of single-layer, free-standing 2D CEY. In particular, we studied the effect strain on the electronic and optical responses of single-layer CEY. 

\begin{figure}[htbp]
\begin{center}
\includegraphics[width=10cm]{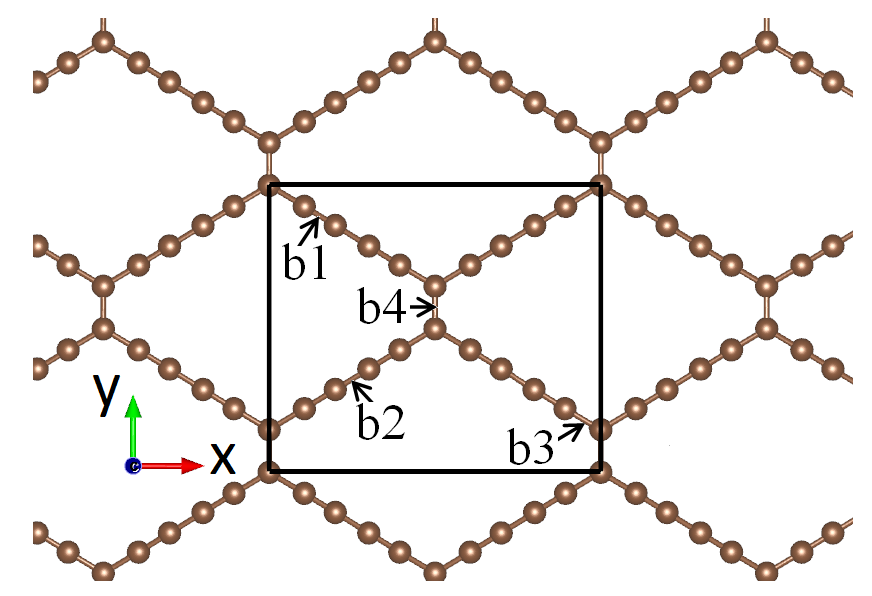}
\caption{Atomic structure of 2D CEY. The black box illustrates the unit-cell used in our calculations. Thermal and optical properties are studied along the x (zizgag) and y (armchair) directions as shown (named in analogy to graphene).}
\label{fig01}
\end{center}
\end{figure}

\section{Methods}

Density functional theory calculations in this study were performed using the Vienna ab-initio simulation package (VASP) \cite{Kresse1996,Kresse1996a,Kresse1999}. We employed the plane wave basis set with an energy cut-off of 500 eV and the exchange-correlation functional proposed by Perdew-Burke-Ernzerhof \cite{Perdew1996}. VMD \cite{Humphrey1996} and VESTA \cite{Momma2011} packages were employed for the illustration of the structures. Fig.\ref{fig01}, illustrates a unit-cell of the CEY lattice, which includes 20 atoms. We studied its properties along the x (zigzag) and y (armchair) directions, as depicted in Fig. \ref{fig01}. The choice of armchair and zigzag nomenclature was chosen in analogy with graphene. The lattice parameters of the geometry optimized structure along the x and y directions were predicted to be 11.26 \AA~ and 9.74 \AA~, respectively. Periodic boundary conditions were applied along all three Cartesian directions such that the obtained results represent the properties of large-area CEY films and not the nanoribbons. A vacuum layer of 20 \AA~ was considered to avoid image-image interactions along the sheet normal direction. 

Once the minimized structure was obtained, we applied loading conditions to evaluate the mechanical response. For this purpose, we increased the length of the periodic simulation box along the loading direction in a multistep procedure, every step with a small engineering strain of 0.001. After changing the simulation box size, the atomic positions were rescaled to avoid any sudden void formation or bond stretching. We then used the conjugate gradient method for geometry optimizations, with termination criteria of $10^{-5}$ eV and 0.01 eV/\AA~ for the energy and the forces, respectively, using a 8x8x1 Monkhorst-Pack \cite{Monkhorst1976} k-point mesh size. 
After the energy minimization process, the stress tensor was calculated to obtain the stress-strain relations. Moreover, the electron localization function for energy minimized structures was simulated by performing a single-step calculation with a denser k-point mesh of 15x15x1. To probe the thermal stability of single-layer, free-standing CEY we performed ab initio molecular dynamics (AIMD) simulations for a 2x2 super-cell with 80 atoms using the Langevin thermostat under constant volume condition, with a time step of 1 fs and 2x2x1 k-point mesh size.

For the electronic and optical properties, the ground state electronic properties were calculated using the PBE functional. Due to the underestimation of the band gap with the PBE functional, we also used the screened hybrid functional, HSE06 \cite{Krukau2006} and quasi-particle many-body perturbation theory (MBPT) via the G$_0$W$_0$ approximation \cite{Shishkin2007} to evaluate the electronic properties of these materials. A 12x12x1 $\Gamma$ centered Monkhorst-Pack k-point mesh was used for the PBE and HSE06 calculations. In the G$_0$W$_0$ approximation, the quasi-particle energies were obtained as a first-order correction to the corresponding Kohn-Sham single-particle energies: {\cite{Shishkin2006,Shishkin2007a}
\begin{equation}
E_{nk}^{QP} = Re[ \langle \phi_{nk}| T+V_{n-e} + V_H + \Sigma_{XC}(G, W, E_{nk})| \phi_{nk} \rangle]
\end{equation}

The calculations were performed with a k-sampling grid of 18x18x1, a vacuum layer of 24 \AA~, 120 empty conduction bands and 200 frequency grid points. Finally, the optical absorption spectra including electron-hole interaction were calculated via the BetheÐSalpeter equation (BSE) \cite{Salpeter1951},
\begin{equation}
(E_{ck}^{QP} - E_{vk}^{QP}) A_{cvk}^S + \sum_{c' v' k'}  \langle cvk | K^{e-h} | c' v' k' \rangle A_{cvk}^S = \Omega^S A_{cvk}^S
\end{equation}
where $A_{cvk}^S$ represents the exciton wavefunction, K$^{e-h}$ is the electron-hole coupling kernel, $\Omega^S$ is the energy of the given excitation, and $E_{ck}^{QP}$ and $E_{vk}^{QP}$ denote the QP eigenvalues of valence and conduction bands at a specific k-point, respectively. In BSE calculations we consider 10 valence and 20 conduction bands with a k-sampling grid of 24x24x1 for all cases.

Classical molecular dynamics modelling in this study was performed using the LAMMPS  package \cite{Plimpton1995}. We employed the Tersoff empirical potential \cite{Tersoff1988} with an optimized parameter set proposed by the Lindsay and Broido \cite{Lindsay2010} to model the interaction between carbon atoms. A recent study has comprehensively compared the available MD forcefields for predicting graphene's thermal conductivity, and it found that the optimized Tersoff potential provides the most accurate predictions\cite{Si2017}. We employed non-equilibrium molecular dynamics (NEMD) simulations to predict the thermal conductivity of CEY at room temperature (300 K). In the NEMD calculations, we used a relatively small time increment of 0.1 fs. Simulations were performed for samples with different lengths to explore the length effect on the predicted thermal conductivity \cite{Mortazavi2017}. For each sample, we applied periodic boundary conditions along the planar directions to remove the effect of free atoms at the edges. We first relax the structures at room temperature using the   {Nos\'e-Hoover thermostat method (NVT).} Before applying the loading condition, we fixed several rows of carbon atoms at the two ends. We then divided the simulation box (excluding the fixed atoms) along the sample length into 20 slabs. The first two and the final two layers were assigned to be either the cold and hot slabs. The temperature gradient was generated by a 20 K temperature difference along the hot and cold layers. During the rest of the simulation, the temperature of these two slabs were controlled at desired values by an NVT thermostat, while the remaining layers  evolved under a constant energy NVE ensemble. We conducted simulations for longer times and averaged the temperatures of every slab. As a result of the temperature gradient, a heat-flux appears in the system, and can be calculated from the energy added to the atoms in the hot slab and the energy removed from the atoms in the cold slab. Based on these energy curves the heat-flux, $q_x$ was calculated. In addition, the averaged temperatures present a linear relation along the sample. The lattice thermal conductivity along each direction $\kappa_{x,y}$, was then evaluated from the one-dimensional form of the Fourier law:
\begin{equation}
\kappa_x = \frac{q_x} {dT/dx}
\end{equation}

\section{Results and discussion}

\subsection{Stretchability of CEY}

To evaluate the response of CEY to mechanical stretch, we applied uniaxial and biaxial loading conditions. For biaxial loading, we applied the strain along both planar directions. In the case of uniaxial loading, we applied the loading strain along the x direction only (as shown in Fig. \ref{fig01}). 
{In order to ensure uniaxial stress condition, the stress along the transverse direction of loading (y direction) should remain around zero. To satisfy this, at every loading strain, the size of the simulation box along the transverse direction was  altered with a goal to reach negligible stress along the transverse direction after the energy minimization of atomic positions using the conjugate gradient method.} 
In Fig. \ref{fig02}, the DFT results for stress-strain responses of single-layer CEY under uniaxial and biaxial strain is illustrated. For biaxial loading, the stress values show  {an initial linear response}, followed by a nonlinear relation up to the ultimate tensile strength point, which occurs at a strain level around 0.15. After the ultimate tensile strength point, stress values drop sharply, consistent with structural failure. In comparison with other 2D materials, the stress-strain response of single-layer CEY under uniaxial loading presents a different pattern. 
 {For pristine graphene under uniaxial loading, the loading directly leads to the increase of carbon-carbon bond lengths oriented along the loading direction and such that the stress value initially increases linearly  \cite{Liu2007,Silvi1994,Gatti2005,Lee2013b}. Meanwhile, the length of carbon-carbon bonds oriented in the transverse direction to the loading shows almost no increase \cite{Rahaman2017}.}
The stress-strain response of CEY nanomembranes under uniaxial strain is non-linear and yields no clear elastic modulus. In this case, up to strain levels of ~0.04, the stress along the loading direction remains almost at zero.
This implies that the bond elongation in the structure is negligible and the deformation was mainly achieved by the contraction along the transverse direction {(see Fig. \ref{fig02} inset)}. After this point, stress values start to increase but smoothly which reveals that the deformation is still mostly achieved by the contraction along the transverse direction, but at the same time the bond elongation is also happening. At strain levels higher than 0.09, the uniaxial stress sharply increases which indicates that the stretching is mainly achieved by bond elongation {(see Fig. \ref{fig02} inset for the different bond lengths)} and such that the sheet contraction along the transverse direction contributes less. At strain levels around 0.27, the structure reaches  {its maximum} load bearing ability which is almost 29 GPa $\cdot$ nm. Shortly after, at strain levels around 0.30, the ultimate tensile strength point is passed and the failure happens in the lattice. For defect-free graphene, the failure strain along the zigzag direction was reported to be 0.194 by Liu et al. \cite{Liu2007}. Because of the porous structure of 2D CEY, it can deform more easily by contraction rather than bond elongation, which explains its considerably higher stretchability in comparison with graphene. Nevertheless, the tensile strength of CEY is approximately  25\% smaller than that of the pristine graphene \cite{Liu2007}.
{The completely non-linear stress-strain response of single-layer CEY under uniaxial loading can be illustrative of its dynamical and elastic instability which is an interesting topic that can be explored  in future studies.}

\begin{figure}[htbp]
\begin{center}
\includegraphics[width=10cm]{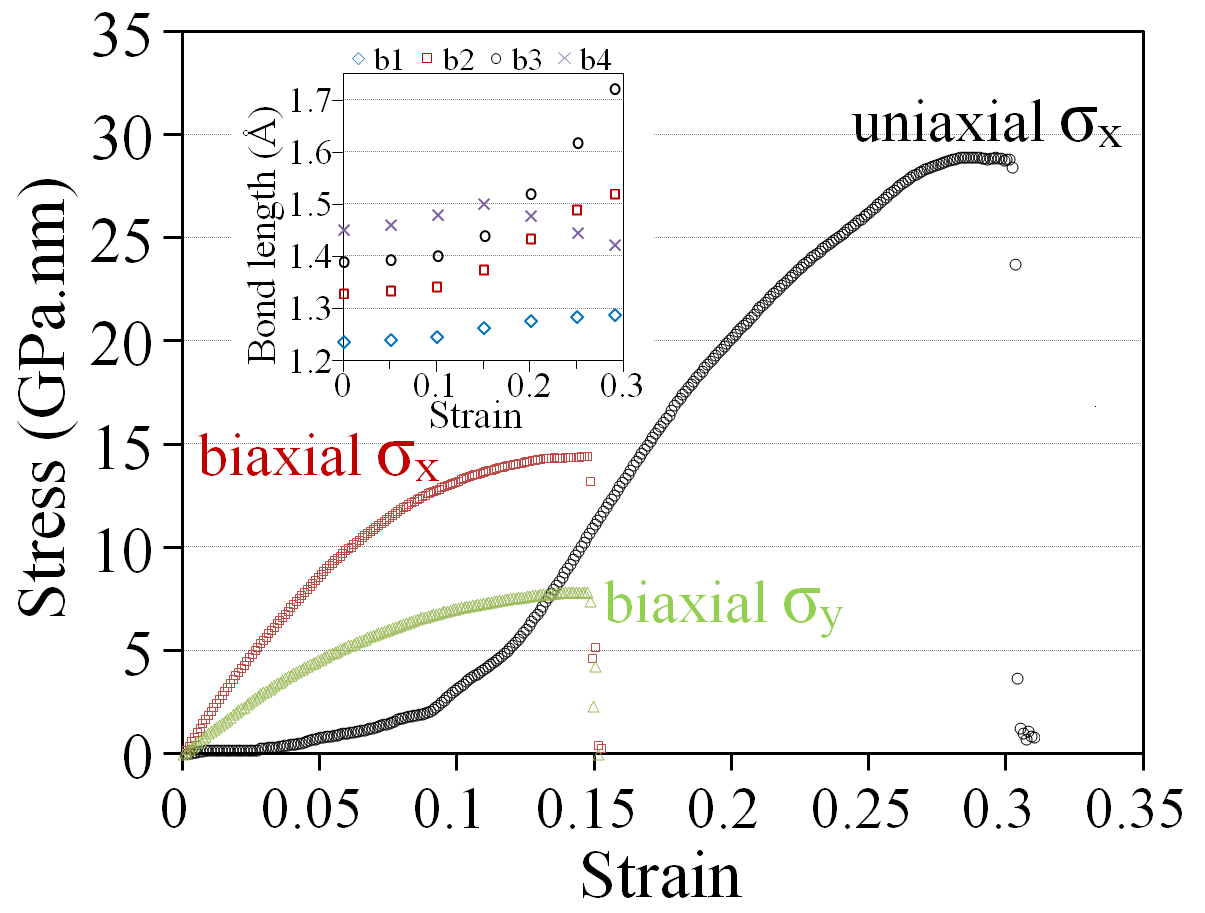}
\caption{Stress-strain responses of single-layer CEY under uniaxial or biaxial loading conditions. 
{Inset compare the evolution of different C-C bond lengths (as illustrated in Fig. \ref{fig01}) during uniaxial loading.}}
\label{fig02}
\end{center}
\end{figure}

To better understand the mechanical response of single-layer CEY under different loading conditions, we analyzed the deformation process as shown in Fig. \ref{fig03}. In this case, in order to correlate electronic structure and mechanical response, we also plotted the electron localization function (ELF) \cite{Silvi1994}. ELF takes a value between 0 and 1, where ELF=1 and ELF = 0.5 correspond to perfect localization and electron gas, respectively. As the first finding, highest electron localization happen at the center of carbon-carbon bonds. The sharing of electrons between two connecting carbon atoms is a characteristic of covalent bonding. The comparison of electron localization on different bonds can be useful to identify their relative strength. In this way, the higher localization of electrons demonstrates the higher rigidity of a particular bond \cite{Mortazavi2017,Gatti2005}. 
{For the structure with minimum energy (Fig. \ref{fig03}a),  sp$^2$-sp$^2$ (b4 bond in Fig. \ref{fig01}) and sp$^2$-sp (b3 bond in Fig. \ref{fig01}) bonds yield the lowest electron localization, respectively. }
Based on our DFT results, electron localization is maximum for the sp-sp (b1 and b2  bonds in Fig. \ref{fig01}) carbon bonds. These preliminary ELF results clearly suggest that the bond stretching and subsequent failure in the CEY structure occur mainly along sp$^2$-sp$^2$ and sp$^2$-sp carbon bonds. For the biaxial loading, the loading is achieved in diagonal direction and in this case the sp-sp  and sp$^2$-sp bonds are more aligned along the loading in comparison with the sp$^2$-sp$^2$ bond. By taking into account the lower rigidity of sp$^2$-sp bonds, we conclude that they stretch more than the other bonds in the lattice (Fig. \ref{fig03}b and Fig. \ref{fig03}c ) and the final rupture happens along these bonds (Fig. \ref{fig03}d). A similar deformation mechanism seem to also take place in  uniaxial loading along the x direction, where the final rupture happens along the sp$^2$-sp carbon bonds (Fig. \ref{fig03}h).  

\begin{figure}[htbp]
\begin{center}
\includegraphics[width=14cm]{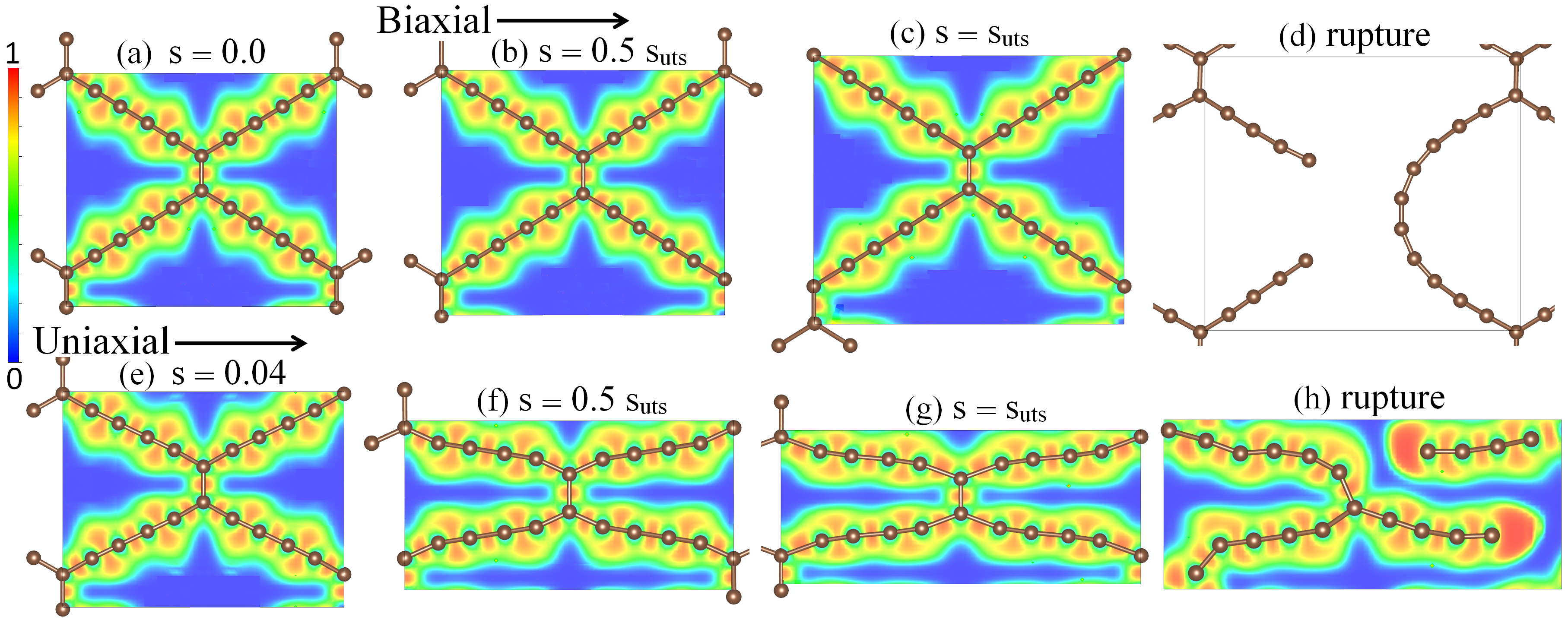}
\caption{Deformation processes of single-layer CEY under uniaxial or biaxial loading conditions at different strain levels (s) with respect to the strain at ultimate tensile strength ($s_{uts}$). The contours illustrate electron localization function (ELF), which takes a value between 0 and 1 (ELF=1 corresponds to perfect localization).}
\label{fig03}
\end{center}
\end{figure}

\subsection{Electronic properties}

To investigate electronic properties of single-layer CEY, the band structure and the total density of states (DOS) were calculated and the obtained results from the PBE, HSE06 and G$_0$W$_0$ were compared. Fig. \ref{fig04} illustrates the band structure and total DOS of energy minimized CEY predicted by the PBE functional. 
{CEY appears to present a very small direct band gap of ~0.04 eV along $\Gamma$-X direction, which is in  excellent agreement with the theoretical PBE result reported in the original study for CEY \cite{Jia2017}.}

\begin{figure}[htbp]
\begin{center}
\includegraphics[width=12cm]{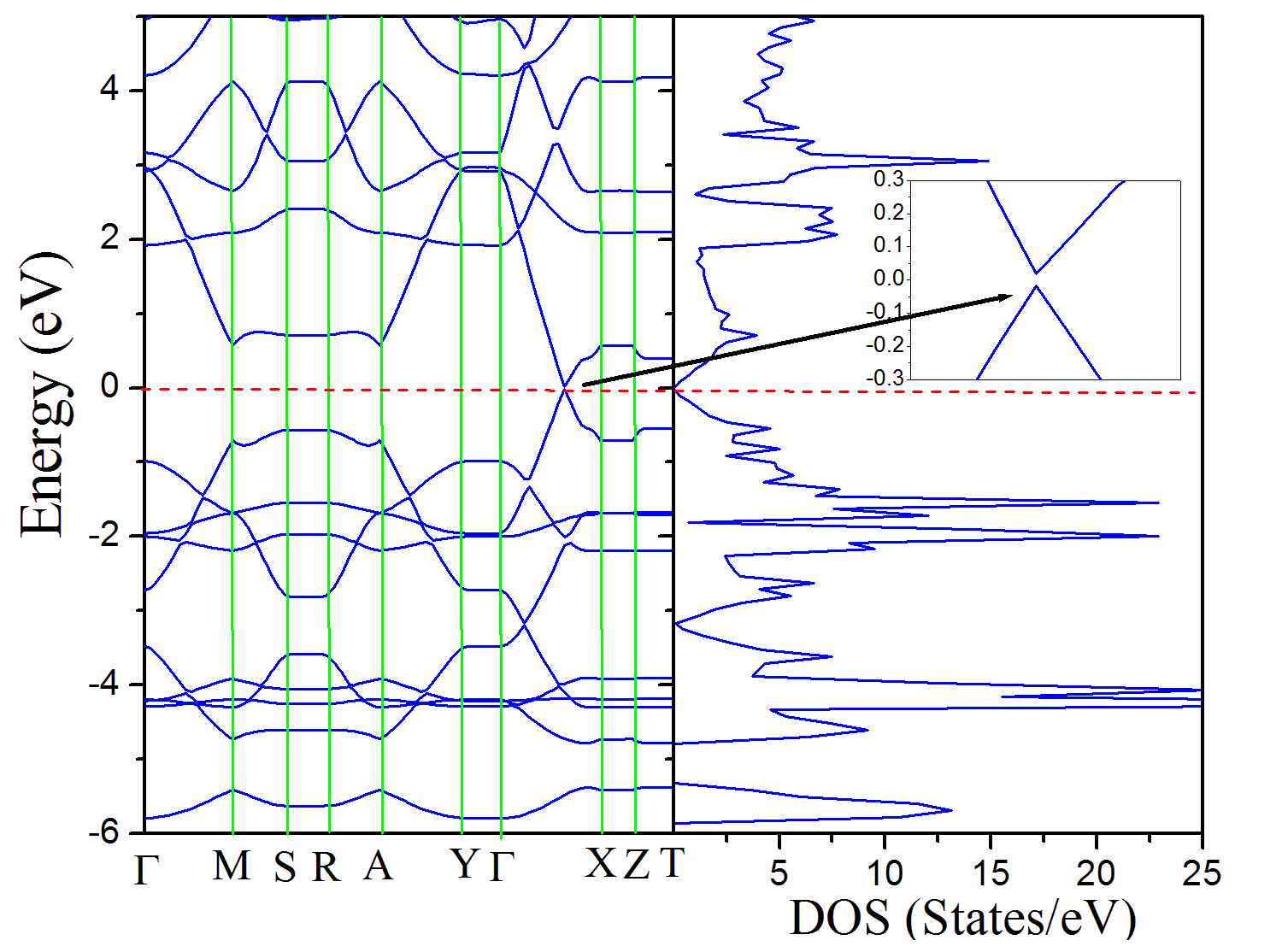}
\caption{Band structure and total DOS of energy minimized single-layer CEY predicted by the PBE functional. Inset shows amplified regions of band structures in the vicinity of the VBM and CBM near the Fermi level.}
\label{fig04}
\end{center}
\end{figure}

In this study, we also explored the evolution of electronic and optical response of single-layer CEY under different loading conditions. To this aim, we selected three different strain levels (s) with respect to the strain at the ultimate tensile strength, s$_{uts}$. In Fig. \ref{fig05}, the band structure and total DOS of strained CEY predicted by PBE are compared. By applying uniaxial or biaxial strain the electronic band gap widens. At both magnitudes of biaxial strain (s=0.5 s$_{uts}$ and s= s$_{uts}$), the valence band maximum (VBM) and the conduction band minimum (CBM) occur along $\Gamma$-X direction, resulting in a direct band gap. The values of band gap in biaxial strained systems for s=0.5 s$_{uts}$ and s=s$_{uts}$ are 0.07 eV and 1.06 eV, respectively. For uniaxial strain, the VBM and CBM are parallel and coincide at Y-$\Gamma$ direction. The corresponding band gap values for these systems are 0.08 (for s=0.5 s$_{uts}$) and 1.25 eV (for s=s$_{uts}$).

\begin{figure}[htbp]
\begin{center}
\includegraphics[width=12cm]{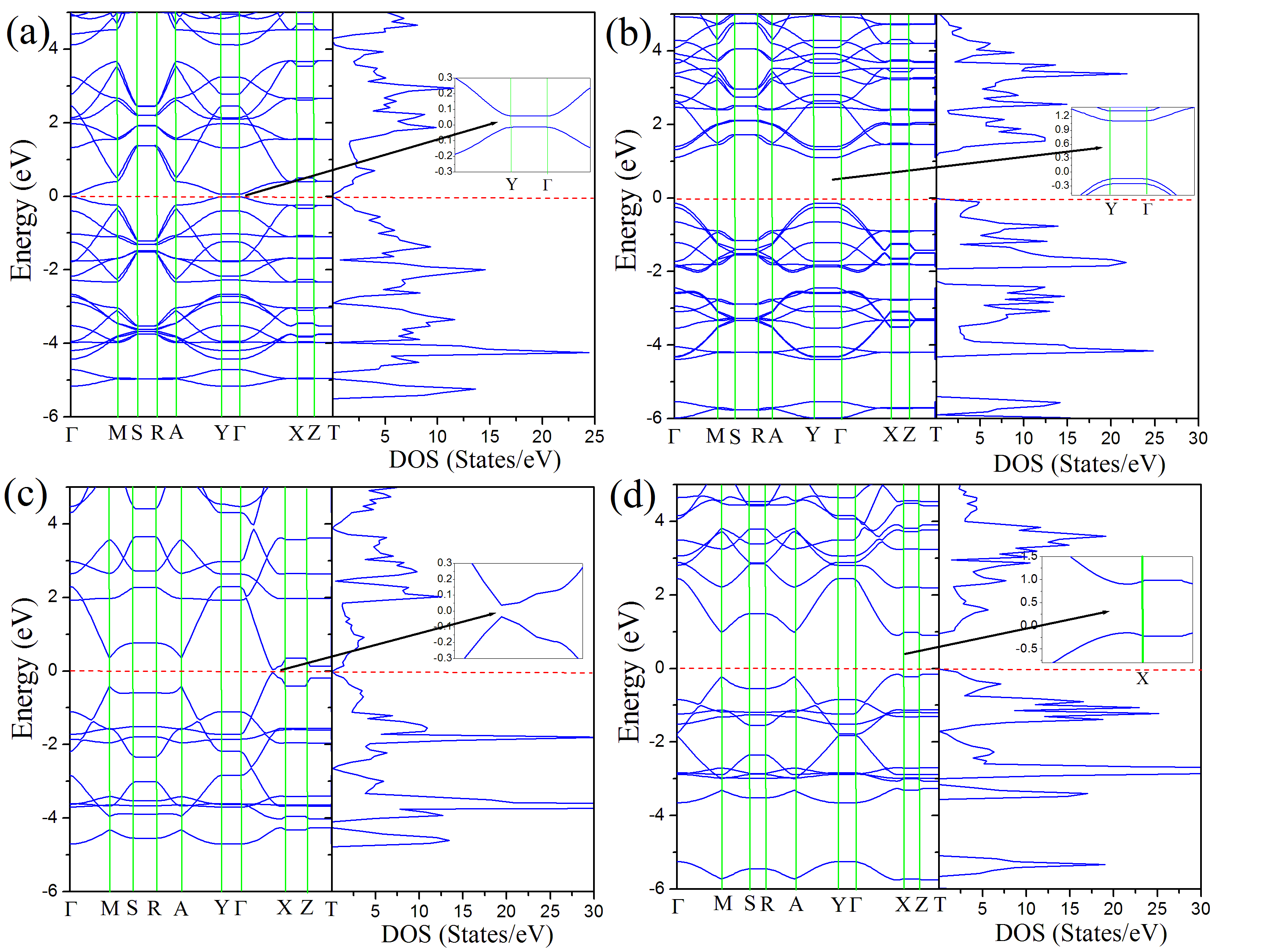}
\caption{Band structure and total DOS of strained CEY predicted by the PBE functional. (a and b) and (c and d) belong to the single layer CEY under uniaxial and biaxial loading conditions, respectively. Here, two different strain levels of (a and c) 0.5 $s_{uts}$ and (b and d) 1 $s_{uts}$ are considered. Insets show amplified regions of band structure in the vicinity of the VBM and CBM near the Fermi level.}
\label{fig05}
\end{center}
\end{figure}

Total DOS for CEY sheets at different strains levels predicted by the HSE06 and G$_0$W$_0$ approaches are shown in Fig. \ref{fig06} (a) and (b), respectively. As expected, the HSE06 functional estimates considerably larger band gaps than the PBE. Table \ref{tab01} summarizes the band gap values for all considered cases. The value of the band gap obtained using HSE06 is 0.54 eV, 0.62 eV and 1.85 eV, respectively, for s=0, s=0.5 s$_{uts}$ and s=s$_{uts}$ for biaxially loaded systems. The corresponding values for uniaxial strained cases are 0.7 eV (for s=0.5 s$_{uts}$) and 2.24 eV (for s=0.5 s$_{uts}$). On the other side, the band gaps for the considered cases obtained by the G$_0$W$_0$ functional are larger than those predicted by the HSE06 functional. For zero strain, the G$_0$W$_0$ method predicts a band gap of 0.88 eV, which is by 0.34 eV larger than that estimated by HSE06. For biaxially strained systems, the G$_0$W$_0$ band gap is 1.02 eV for s= 0.5 s$_{uts}$ and 2.35 eV for s=s$_{uts}$. For the uniaxially strained CEY films, the G$_0$W$_0$ functional predicts band gaps of 1.07 eV for s= 0.5 s$_{uts}$ and 2.6 eV for s= s$_{uts}$. It can be concluded that by applying uniaxial or biaxial strains on this newly synthesized allotrope of carbon, the band gap can be continuously tuned. In general, the band gaps of uniaxially strained single-layer CEY are larger than those biaxially loaded as evaluated by PBE, HSE06 and G$_0$W$_0$ methods. 
 
 \begin{figure}[htbp]
\begin{center}
\includegraphics[width=14cm]{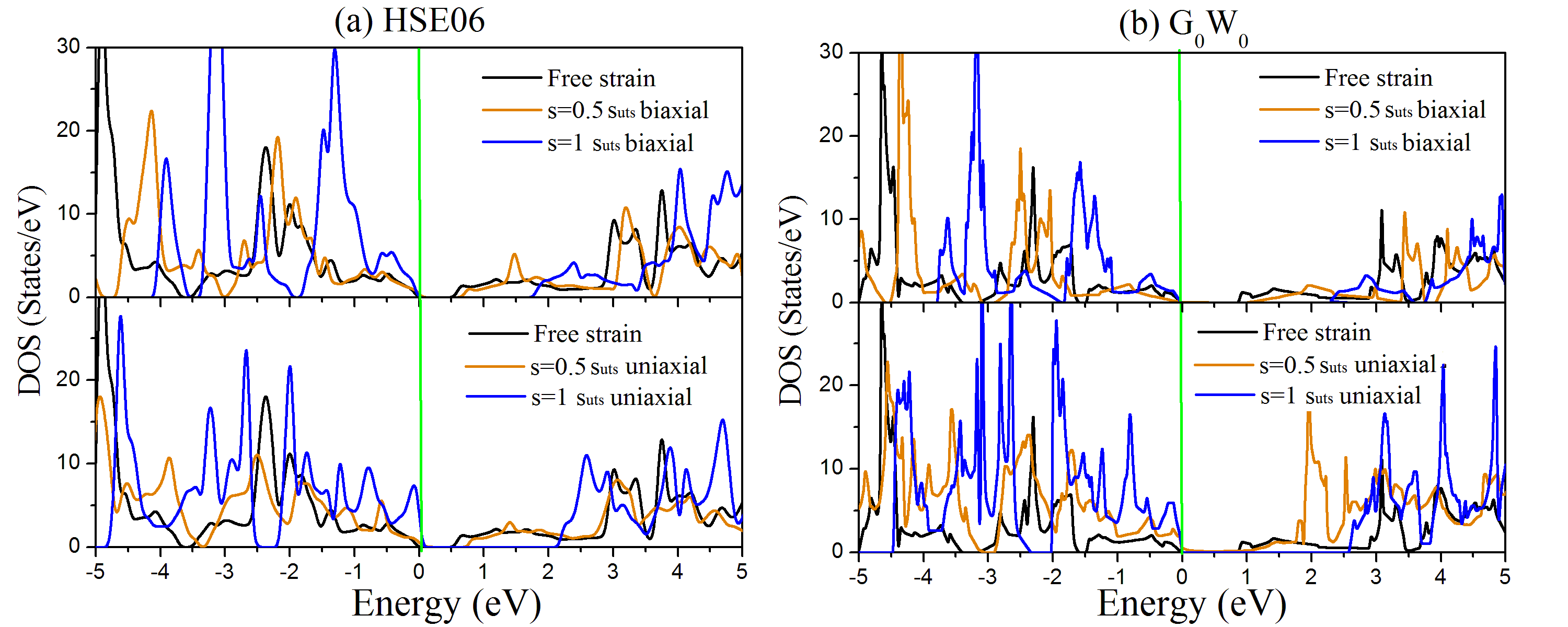}
\caption{Evolution of total density of states for pristine and single-layer CEY under different loading conditions and strain levels, predicted by (a) HSE06 and (b) G$_0$W$_0$ approaches. The Fermi energy is aligned to zero.}
\label{fig06}
\end{center}
\end{figure}

\begin{table}[htp]
\caption{The electronic band gap of single-layer CEY predicted by PBE, HSE06 and G$_0$W$_0$ approaches (the units are in eV).}
\begin{center}
\begin{tabular}{|c|c|c|c|}
\hline
Structure & PBE & HSE06 & G$_0$W$_0$ \\
\hline
s=0 & 0.04 & 0.54 & 0.88 \\
s=0.5 s$_{uts}$, uniaxial & 0.08 & 0.70 & 1.07 \\
s= s$_{uts}$, uniaxial & 1.25 & 2.24 & 2.6 \\
s=0.5 s$_{uts}$, biaxial & 0.07 & 0.62 & 1.02 \\
s= s$_{uts}$, biaxial & 1.06 & 1.85 & 2.35 \\
\hline
\end{tabular}
\end{center}
\label{tab01}
\end{table}%

\subsection{Optical properties}
 
In this section we apply the BSE method to investigate the optical properties of CEY.  Because of the strong depolarization effect in the 2D planar geometry for light polarization perpendicular to the plane, only the optical properties for light polarization parallel to the plane are reported. Moreover, due to the asymmetric geometry of these systems along the x- and y-axes, their optical spectra are anisotropic for light polarizations along the x-axis (E$\vert \vert$x) and y-axis (E$\vert \vert$y). Hence, the optical properties for both parallel polarizations (E$\vert \vert$x and E$\vert \vert$y) are reported. The imaginary part of the dielectric function versus photon energy in unstrained and strained systems at different magnitudes of strain for E$\vert \vert$x and E$\vert \vert$y obtained using G$_0$W$_0$+BSE are presented in Fig. \ref{fig07}. The anisotropic response for different polarizations can be observed for these nanomembranes. The lowest optically active (bright) exciton in the unstrained case was observed at 0.84 eV (0.75 eV) in E$\vert \vert$x (E$\vert \vert$y), which presents a binding energy of 0.04 eV (0.11 eV) and is due to direct interband transitions involving the VBM and CBM. Table \ref{tab02} summarizes the lowest three bright excitons and binding energies  for all nanostructures along x- and y-axis. We observe that when strain is applied, the first three exciton peaks are shifted to higher energies (blue shift). The first excitonic peaks appear almost at the same energies (below 1 eV) for systems under low strain, while under high strain these peaks occur over 2 eV. The positions of the first exciton peaks were observed at 1.0 eV (0.91 eV) and 2.32 eV (2.15 eV) in E$\vert \vert$x (E$\vert \vert$y) for s= 0.5 s$_{uts}$ and s= s$_{uts}$ of uniaxially strained CEY, respectively, which becomes photoactivated along Y-$\Gamma$ wavevectors. The corresponding values in biaxially strained structures are 0.93 eV (0.84eV) and 2.03 eV (1.85 eV) for s= 0.5 s$_{uts}$ and s= s$_{uts}$, respectively. 

 \begin{figure}[htbp]
\begin{center}
\includegraphics[width=14cm]{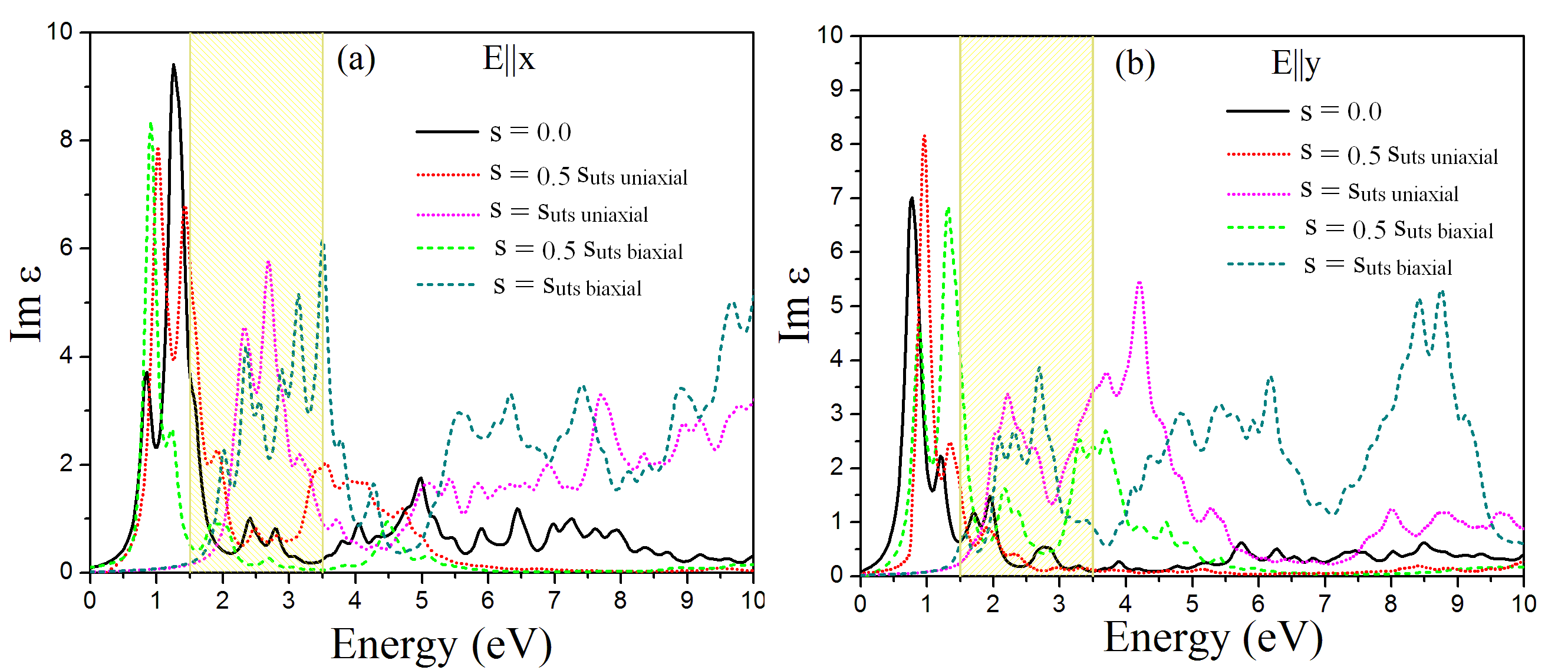}
\caption{Imaginary part of the dielectric function versus photon energy at different magnitudes of strain for light polarizations (a) parallel to the x-axis (E$\vert \vert$x), (b) parallel to the y-axis (E$\vert \vert$y). Shaded area shows the visible range of the spectrum.}
\label{fig07}
\end{center}
\end{figure}

\begin{table}[htp]
\caption{The first (E$_1$), second (E$_2$), third (E$_3$) exciton energies and lowest exciton binding energy (E$_B$) calculated by G$_0$W$_0$-BSE of unstrained and strained systems. The values outside and inside of parenthesis belong to E$\vert \vert$x and E$\vert \vert$y, respectively, with units are in eV.}
\begin{center}
\begin{tabular}{|c|c|c|c|c|}
\hline
Structure & E$_1$ & E$_2$ & E$_3$ & E$_B$ \\
\hline
s=0 & 0.84 (0.77) & 1.26 (1.21) &1.57 (1.71) & 0.04 (0.11)  \\
s=0.5 s$_{uts}$, uniaxial & 1.0 (0.91) & 1.43 (1.35) & 1.93(1.91) & 0.07(0.16) \\
s= s$_{uts}$, uniaxial & 2.32 (2.0) & 2.68 (2.21) & 2.9 (2.42) & 0.28 (0.60) \\
s=0.5 s$_{uts}$, biaxial & 0.93 (0.84) & 1.22 (1.31) & 1.8 (2.17) & 0.09 (0.18) \\
s= s$_{uts}$, biaxial & 2.03 (1.60) & 2.34 (2.10) & 2.56 (2.31) & 0.30 (0.75) \\
\hline
\end{tabular}
\end{center}
\label{tab02}
\end{table}%

The blue shift can also be seen for the second and third exciton peaks in strained systems.  The exciton energies reveal that the third exciton peaks for unstrained and low strain systems are located in visible range of the spectrum. Meanwhile, for systems under maximum strain the first excited-state is also in the visible range. These results indicate that these systems can absorb visible light. In order to better understand the optical properties, the binding energy of the lowest energy exciton (E$_B$) has been calculated as E$_B$ = E$_g$ - E$_{ex}$, where E$_g$g is the quasi-particle band gap from G$_0$W$_0$ calculations and E$_{ex}$ is the excitation energy. From Table \ref{tab02}, we concluded that by increasing the magnitude of strain the binding energy of these systems increases. The binding energies for systems under lower strain levels are larger than those for excitons in Si and GaAs, corresponding to 14.7 meV and 4.2 meV. Whereas, for maximum strain cases they are close to graphene oxide 0.35-0.5 eV \cite{Lee2013b}, graphydine 0.55 eV \cite{Luo2011}, ZnS monolayer 0.36 eV \cite{Shahrokhi2016}, Si-doped graphene 0.22-71 eV \cite{Shahrokhi2017}, and organic electroluminescent (EL) materials, e.g.  polydiacetylene (PDA)  0.5  eV,  polythiophene  (PT)  0.5  eV,  CuPc  0.6  eV \cite{Knupfer2003}. The exciton binding energies for these systems are smaller than those found for wide-band gap semiconductors and insulators such as BN sheets 2.1 eV, graphane 1.6 eV and BeO nanosheet \cite{Shahrokhi2016a}. The optical properties from our BSE calculations raise the possibility of using 2D CEY in optoelectronics and electronics. Moreover, these materials show anisotropic optical responses for E$\vert \vert$x and E$\vert \vert$y, which might be promising as a good optical linear polarizer. 

We next discuss the optical conductivity of 2D CEY. The real part of the optical conductivity is related to the $Im[\epsilon_{\alpha \beta} (\omega)]$ and is given by \cite{Moradian2012}:
\begin{equation}
Re [\kappa_{\alpha \beta}(\omega)] = \frac{\omega}{4 \pi} Im [\epsilon_{\alpha \beta} (\omega)]
\end{equation}
The optical conductivity of unstrained and strained 2D CEY are compared in Fig. \ref{fig08}. It is found that the conductivity for all systems starts with a gap which indicates the semiconducting character of these structures. The optical conductivity for maximum strained systems below 2 eV is zero, where it is maximum for the systems under lower strain. The optical conductivity for the systems under maximum strain increases at energy range over 2 eV by more than 2 times. We have also compared the optical conductivity of these materials with pristine graphene in the visible light range (from 350 to 750 nm) as a function of the wavelength. The insets in Fig. \ref{fig08} show the optical conductivity of aforementioned 2D structures using the G$_0$W$_0$-BSE method in the visible range. 
{We have also included our previous theoretical results for graphene \cite{Shahrokhi2017} obtained using G$_0$W$_0$+BSE method for comparison in this figure. As we showed previously \cite{Rahaman2017}, the optical absorption spectra and the electron energy-loss spectrum of pristine graphene are in a good agreement with experimental values \cite{Shahrokhi2017,Nair2010}. The many-body GWA approach provides a correct description of the electronic structure and leads to more accurate band gaps in semiconductors. Moreover, the electron-hole two-particle Green function in Bethe-Salpeter equation approach (BSE) allows to develop the optical response of a system. So, we believe that the present results of G$_0$W$_0$ band gaps and optical properties from G$_0$W$_0$+BSE method are completely reliable and should well compare with eventual experiment.}
It can be seen that the optical conductivity for CEY under  lower strain in both polarizations is smaller than that of graphene. Only at wavelength range between 600 and 670 nm (orange and red), the optical conductivity of lower strained systems is a little larger than the graphene one. At wavelength range between 350 and 650 nm the optical conductivity for the systems under maximum strain is larger than graphene. This enhancement in optical conductivity for the maximum strained systems suggest them as promising candidates to be used in photovoltaic cells.

 \begin{figure}[htbp]
\begin{center}
\includegraphics[width=14cm]{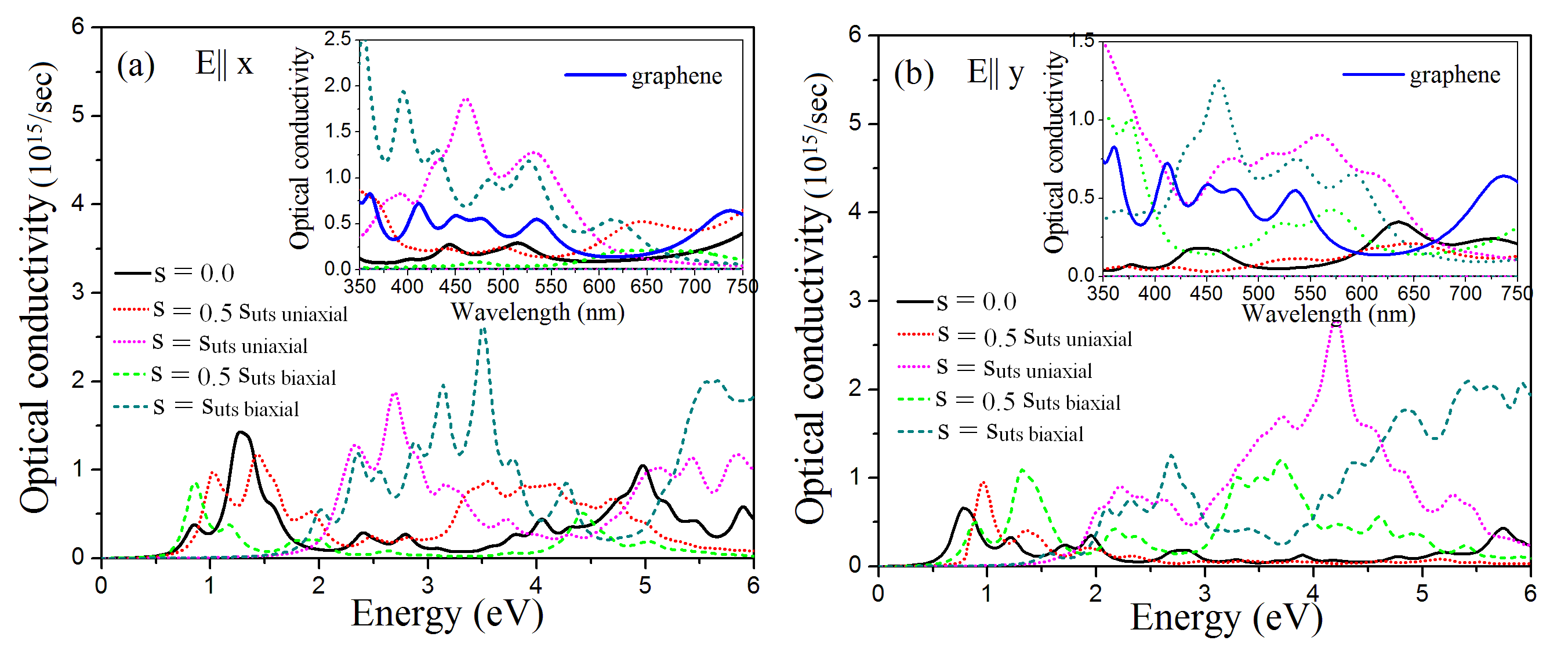}
\caption{The optical conductivity as a function of photon energy at different magnitudes of strain for light polarizations (a) parallel to the x-axis (E$\vert \vert$x), (b) parallel to the y-axis (E$\vert \vert$y). Insets show a comparison of optical conductivity as a function of wavelength, for aforementioned materials and graphene in the visible range (350-750 nm). Our previous theoretical results for graphene obtained using G$_0$W$_0$+BSE are also reported in this figure [46].}
\label{fig08}
\end{center}
\end{figure}

\subsection{Stability at finite temperatures}

All the calculations done so far present results which assume a temperature of 0 K. 
{Next, we study the thermal stability of single-layer CEY by performing AIMD simulations, which have been widely used for predicting the thermal stability of 2D materials \cite{Broek2014}. In this case, we conducted the AIMD simulations at different temperatures for a time of 15 ps. As shown in Fig. \ref{fig09}, single-layer CEY remains stable at a high temperature of 1500 K. Interestingly, our simulation results for the higher temperatures up to 3000 K does not reveal any sign of evaporation of CEY and instead the structures convert to long carbon chains. As it can be observed in the Fig. \ref{fig09}, the debonding happen in the sp$^2$-sp$^2$ carbon bonds. This observation is in agreement with our ELF analysis which revealed the lowest rigidity for these carbon bonds. The high thermal stability of CEY up to 1500 K confirms also its suitability for the applications at temperatures higher than  ambient conditions.}

 \begin{figure}[htbp]
\begin{center}
\includegraphics[width=14cm]{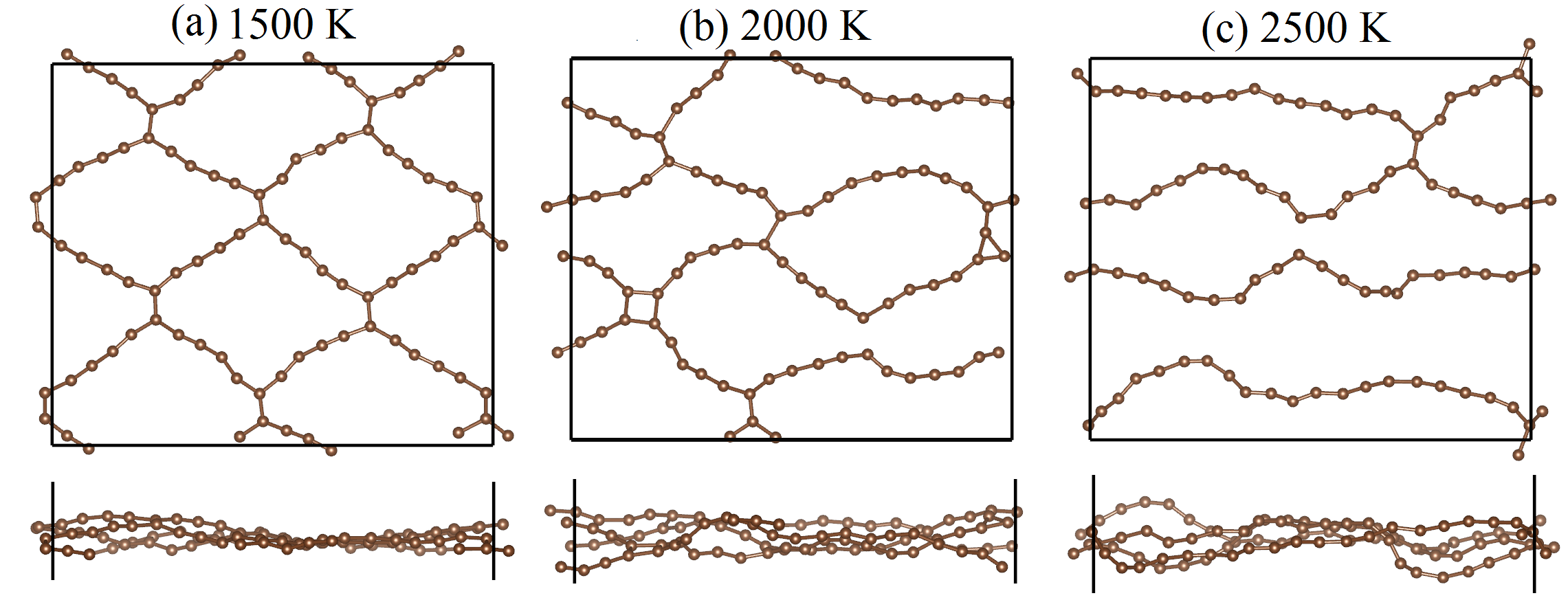}
\caption{Top and side snapshots of single-layer CEY structure after AIMD simulations for 15 ps at different temperatures.}
\label{fig09}
\end{center}
\end{figure}

\subsection{Thermal conductivity}

Finally, we explore the lattice thermal conductivity of single-layer CEY using classical NEMD simulations. In this case, to ensure the length independency of the predicted thermal conductivity, we carried out NEMD simulations for CEY samples of increasing  length. In Fig. \ref{fig10}, the NEMD results for the thermal conductivity of single-layer CEY along the x- and y- directions at room temperature are illustrated. A slight increase in the thermal conductivity is observable for the samples with lengths below 60 nm. However, for the CEY samples longer than 60 nm the thermal conductivity along both x- and y- directions converges which verifies that the thermal transport becomes size independent and it is within the diffusive heat conduction regime. 
The relation between the thermal conductivity in finite systems, $\kappa(L)$ and its intrinsic value is given by { \cite{Schelling2002,Pereira2016}}
\begin{equation}
\frac{1}{\kappa(L)} = \frac{1}{\kappa} \left( 1 + \frac{\Lambda_{eff}}{L} \right),
\end{equation}
where $\Lambda_{eff}$ is an effective phonon mean free path (MFP).
Adjusting the above equation to our NEMD results, the length-independent phononic thermal conductivity of single-layer CEY at room temperature along the x- and y- directions are predicted to be 10.2 $\pm$ 0.9 W/mK and 3.44 $\pm$ 0.30 W/mK, respectively. 
{Meanwhile, the effective phonon mean free paths are 4.2 $\pm$ 0.2 nm and 11 $\pm$ 0.5 nm, along x- and y- directions respectively. It is interesting to notice that the direction with lower $\kappa$ has a higher $\Lambda_{eff}$. This reflects the nature of $\Lambda_{eff}$, which is just an average over all modes with very different MFP. Therefore, $\Lambda_{eff}$ might not be suitable for comparing the thermal conductivity along each direction.}
These findings reveal low and highly anisotropic phonon thermal transport along 2D CEY. The thermal conductivities of single-layer CEY are almost three orders of magnitude smaller than that of pristine graphene \cite{Balandin2008,Ghosh2010,Balandin2011,Xu2014,Fan2017}.

 \begin{figure}[htb]
\begin{center}
\includegraphics[width=12cm]{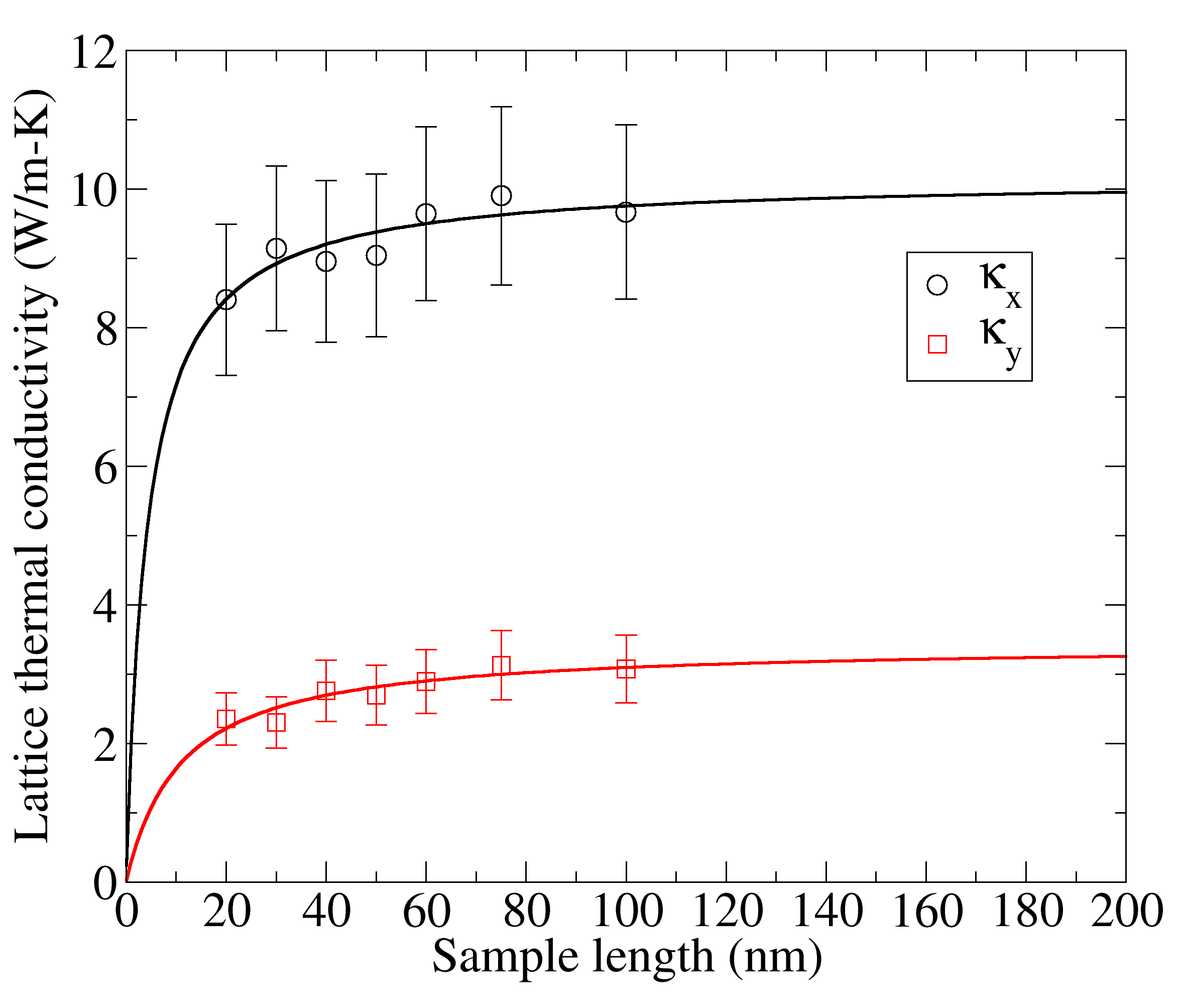}
\caption{Non-equilibrium classical molecular dynamics results for the length effect on the thermal conductivity of single-layer CEY along the x- and y- directions at the room temperature. We assumed a thickness of 3.35 \AA, the same as graphene.}
\label{fig10}
\end{center}
\end{figure}

In order to evaluate the anisotropy observed in $\kappa$ for CEY and as a means to compare with results for graphene, we have calculated the phonon group velocities from phonon dispersion along x- and y- directions and compared it to graphene, as shown in Fig. \ref{fig11}. From the figure we can infer that the group velocities in CEY along both directions are much smaller than the group velocities for graphene. Furthermore, if we consider the average group velocities, we have averages of 115 m/s for the x-direction of CEY and 71 m/s for the y-direction of CEY. So, the direction with lower thermal conductivity also presents lower group velocities. For comparison, the average group velocity in graphene is 740 m/s independent of direction. So, our results are consistent with a lower thermal conductivity for CEY relative to graphene.

 \begin{figure}[htb]
\begin{center}
\includegraphics[width=12cm]{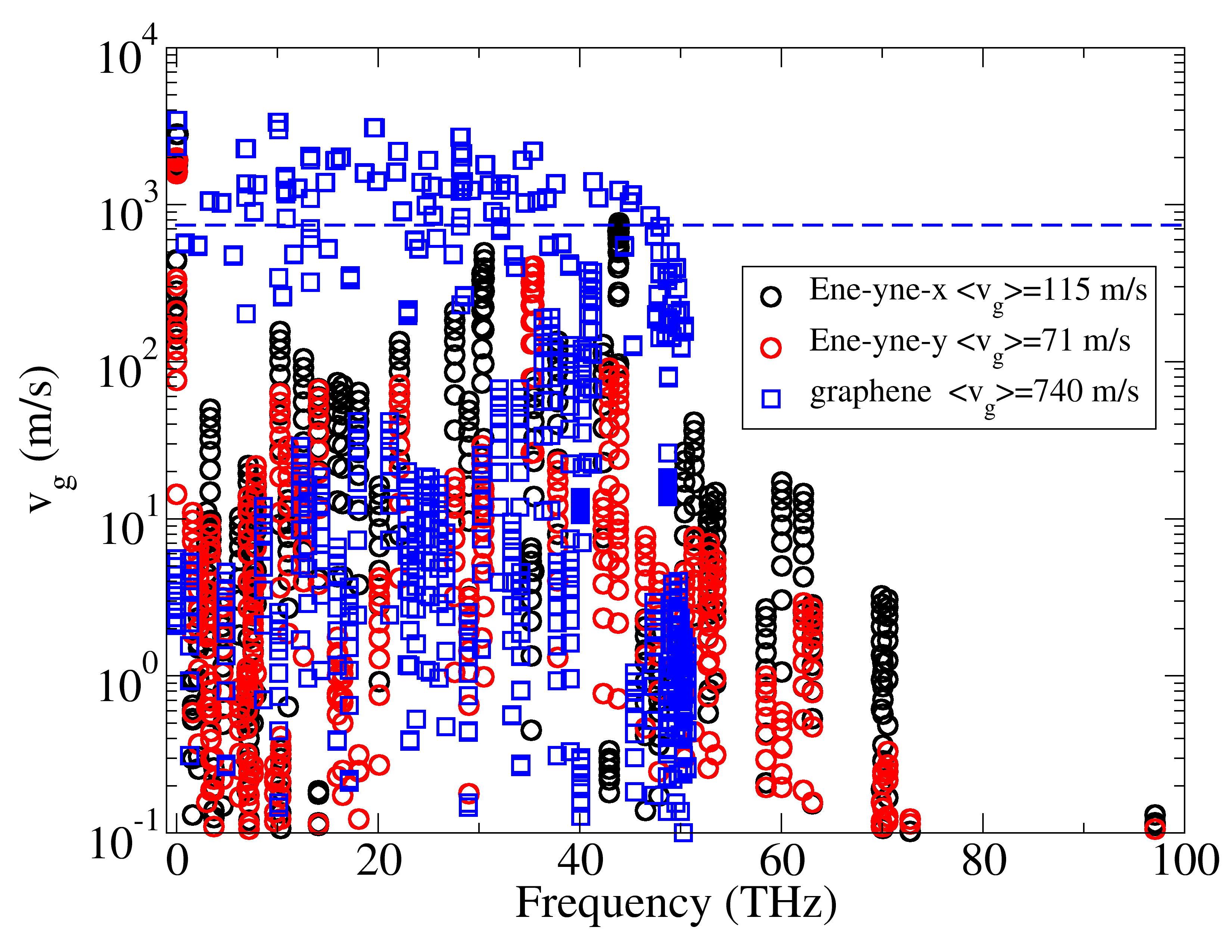}
\caption{Phonon group velocities calculated for CEY along x- and y-directions and comparison to graphene. Group velocities are much smaller for CEY, which is consistent with its lower thermal conductivity.}
\label{fig11}
\end{center}
\end{figure}

\section{Conclusions}

Three decades after its theoretical prediction, carbon Ene-yne (CEY), a novel carbon-based material was fabricated.  Motivated by this recent experimental advance, we conducted extensive first-principles calculations and classical MD simulations to explore the elasticity, electronic structure, optical response, and lattice thermal conductivity of free-standing and single-layer CEY. Because of the porous structure of CEY, it was found that it can deform easily by contraction in the transverse direction rather than by bond elongation, such that CEY presents ~50\% higher stretchability in comparison with pristine graphene. Remarkably, the tensile strength of CEY was predicted to be as high as 29 GPa.nm, which is only 25\% smaller than that of the pristine graphene. We also  explored the change in electronic structure and optical properties of this newly synthesized 2D carbon allotrope under uniaxial or biaxial strain within the framework of density functional theory, using the PBE and the HSE06 functional, the G$_0$W$_0$ method and the BetheÐSalpeter equation. 
{Our first-principles results confirm that CEY present semiconducting electronic character with a small band gap, very promising for applications in stretchable and flexible nanoelectronics. The values of band gap for CEY obtained using PBE, HSE06 and G$_0$W$_0$ are 0.04, 0.54 and 0.88 eV, respectively.}
We also found that by applying an external strain, the band gap of this system can be finely tuned. The imaginary part of the dielectric function, exciton binding energies and optical conductivity of the material were investigated by the BSE+G$_0$W$_0$ method. We found that by applying strain on this allotrope, the first three exciton peaks are shifted to higher energies (blue shift). Moreover, the optical conductivity for the structures under strain  close to the tensile strength point, is larger than for graphene in the visible range. The effects on electronic structure and enhancement in optical conductivity of these systems are potentially useful for novel optoelectronic nanodevices. Furthermore, single-layer CEY show anisotropic optical responses for E$\vert \vert$x and E$\vert \vert$y, which are attractive for applications in optical linear polarizers. Ab-initio molecular dynamics simulations show that single-layer CEY can endure temperatures up to 1500 K, suitable for high temperature application. According to our classical non-equilibrium molecular dynamics simulations, the thermal conductivity of single-layer CEY at room temperature was found to be anisotropic and almost three orders of magnitude smaller than that of the graphene, 10.2 $\pm$ 0.9 W/mK and 3.44 $\pm$ 0.30 W/mK, along the zigzag and armchair directions, respectively. The phonon group velocities are lower along the y-direction, in agreement with the expected thermal conductivities. {The low thermal conductivity along with direct band gap semiconducting electronic character of CEY might be appealing for the thermoelectric applications. }
Our study provides a comprehensive insight in the mechanical, electronic, optical and thermal transport properties of CEY and show that this new 2D carbon allotrope has potential for application in stretchable and flexible nanodevices.

\section*{Acknowledgments}
B.M. and T.R. greatly acknowledge the financial support by European Research Council for COMBAT project (Grant no. 615132).
L.F.C.P. acknowledges financial support from Brazilian government agency CAPES for project ``Physical properties of nanostructured materials" via its Science Without Borders program (Grant no. 3195/2014).

\end{document}